\documentclass[aps,pra,twocolumn,superscriptaddress]{revtex4-1}

\usepackage[utf8]{inputenc}
\usepackage[english]{babel}
\usepackage[T1]{fontenc}

\usepackage{mathtools}
\usepackage{amsmath}
\usepackage{bm}
\usepackage{stmaryrd}
\usepackage{braket}
\usepackage{amsfonts}
\usepackage{subcaption}
\usepackage{dcolumn}
\usepackage{graphicx}
\usepackage[colorlinks=true, allcolors=blue]{hyperref}
\hypersetup{
  colorlinks   = true, 
  urlcolor     = blue, 
  linkcolor    = blue, 
  citecolor   = blue 
}
\usepackage[justification=justified]{caption} 
\usepackage{ulem}
\renewcommand{\emph}{\textit}
\usepackage{ragged2e}
\DeclareCaptionJustification{justified}{\justifying}

\begin{document}

\title{Real-Time Source Independent Quantum\\Random Number Generator with Squeezed States}

\author{Thibault Michel}
\email{thibault.michel@lkb.upmc.fr}
\affiliation{Center for Quantum Computation and Communication Technology, Department of Quantum Science, The Australian National University, Canberra, ACT 0200, Australia}
\affiliation{Laboratoire Kastler Brossel,
UPMC-Sorbonne Universit\'es, CNRS, ENS-PSL Research University, Coll\`ege de France,
4 place Jussieu, 75252 Paris, France}
\author{Jing Yan Haw}
\affiliation{Center for Quantum Computation and Communication Technology, Department of Quantum Science, The Australian National University, Canberra, ACT 0200, Australia}
\author{Davide G. Marangon}
\affiliation{Dipartimento di Ingegneria dell’Informazione, Università degli Studi di Padova, Via Gradenigo 6B, 35131 Padova, Italy}
\author{Oliver Thearle}
\affiliation{Center for Quantum Computation and Communication Technology, Department of Quantum Science, The Australian National University, Canberra, ACT 0200, Australia}
\author{Giuseppe Vallone}
\affiliation{Dipartimento di Ingegneria dell’Informazione, Università degli Studi di Padova, Via Gradenigo 6B, 35131 Padova, Italy}
\affiliation{Istituto di Fotonica e Nanotecnologie
—CNR, Via Trasea 7, 35131 Padova, Italy}
\author{Paolo Villoresi}
\affiliation{Dipartimento di Ingegneria dell’Informazione, Università degli Studi di Padova, Via Gradenigo 6B, 35131 Padova, Italy}
\affiliation{Istituto di Fotonica e Nanotecnologie
—CNR, Via Trasea 7, 35131 Padova, Italy}
\author{Ping Koy Lam}
\email{ping.lam@anu.edu.au}
\affiliation{Center for Quantum Computation and Communication Technology, Department of Quantum Science, The Australian National University, Canberra, ACT 0200, Australia}
\author{Syed M. Assad}
\affiliation{Center for Quantum Computation and Communication Technology, Department of Quantum Science, The Australian National University, Canberra, ACT 0200, Australia}

\begin{abstract}
  Random numbers are a fundamental ingredient for many applications
  including simulation, modelling and cryptography. Sound random
  numbers should be independent and uniformly distributed. Moreover,
  for cryptographic applications they should also be unpredictable. 
  We demonstrate a real-time self-testing source independent quantum random number generator (QRNG) that uses squeezed light as source. We generate secure random numbers by measuring the quadratures of the electromagnetic field
  without making any assumptions on the source; only the detection
  device is trusted. We use a homodyne detection to alternatively
  measure the $\hat{Q}$ and $\hat{P}$ conjugate quadratures of our
  source. Using the entropic uncertainty relation, measurements on
  $\hat{P}$ allows us to estimate a bound on the min-entropy of
  $\hat{Q}$ conditioned on any classical or quantum side information
  that a malicious eavesdropper may detain. This bound gives the
  minimum number of secure bits we can extract from the $\hat{Q}$
  measurement. We discuss the performance of different estimators for this bound. We operate this QRNG with a squeezed state and we compare its performance with a QRNG using thermal states.  The
  real-time bit rate was $8.2$ kb/s when using the squeezed
  source and between $5.2$--$7.2$ kb/s when the thermal state source
  was used.
\end{abstract}

\maketitle
\section{Introduction}

Random numbers are used as a resource for many application such as
statistical analysis, numerical simulation, encryption and
communication protocols. Random numbers must satisfy three main
requirements: they must be uniformly distributed, independent and
unpredictable.  Pseudo random numbers are generated with a computer
via algorithmic routines from a seed. They have the advantage of being
easy to implement and fast, but they are intrinsically not secure due
to their deterministic generation~\cite{hellekalek_good_1998} and some
commonly used pseudo random number generators (PRNG) have been shown
to be unsecure~\cite{dodis_security_2013}. Their randomness can also
be flawed~\cite{herrero-collantes_quantum_2017} which can lead to
errors in simulations~\cite{Marsaglia68,
  ferrenberg_monte_1992}. Physical random number generators use a
stochastic physical process as the source of
randomness~\cite{uchida_fast_2008, marangon_random_2015}. They are
slower than PRNGs but can still achieve very high generation rate and
has been used as a seed for PRNGs. For random number generators based
on classical systems, the randomness usually originate as lack of
knowledge on the initial state of the system, in which case the
security relies on the assumption that no one has a better knowledge
of this original state.  Quantum systems on the other
hand~\cite{ma_quantum_2016} offer an interesting alternative source of
randomness as measurement outcomes on such systems are intrinsically random due to
Born's rule~\cite{rarity1994quantum}. It is then possible to create a
long-term stable \cite{marangon_long-term_2018}, fast
quantum random number generator (QRNG)~\cite{symul_real_2011,haw_maximization_2015,zhang_quantum_2017},
in a self testing fashion~\cite{lunghi_self-testing_2015}, or even on
a mobile phone~\cite{sanguinetti_quantum_2014}. However the
measurement outcomes may still be correlated with another
party~\cite{frauchiger_true_2013}. This is the case whenever the
source of randomness is in a mixed state. Even in that case, it is
still possible to exploit non-local Bell state
measurement~\cite{bell1964einstein, brunner_bell_2014} to extract true
random numbers without any assumption on the source of randomness or
the measurement
device~\cite{pironio_random_2010,christensen2013detection,pivoluska2014device,liu2018high,bierhorst_experimentally_2018}.
However these implementation are still very slow with rates around few
tens of bits per second.  In a similar fashion, generation protocols
using light emitted from distant cosmic sources were recently proposed
and
demonstrated~\cite{wu2017random,handsteiner2017cosmic,leung2018astronomical}. As
a faster alternative one can implement a semi device independent QRNG
by assuming only either the source~\cite{nie_experimental_2016} or the
detection~\cite{vallone_quantum_2014,xu_high_2017,
  marangon_source-device-independent_2017,ma_source-independent_2017,
  avesani_source-device-independent_2018} device is trusted. In a
source independent quantum random number generator (SI-QRNG), the
source of randomness can be arbitrary and controlled by an adversarial
party; yet it can still yield secure random numbers. Roughly speaking,
the principle of SI-QRNG is that, by switching between different
measurement basis, one is able to assess the purity of the source, which
can in turn set a bound on its extractable randomness. This can be
formalized rigorously using the entropic uncertainty
relation~\cite{furrer_position-momentum_2014} which was first
introduced in~\cite{bialynicki-birula_uncertainty_1975}.

SI-QRNG based on the entropic uncertainty relation have already been demonstrated in both discrete~\cite{vallone_quantum_2014}
and continuous variables~\cite{marangon_source-device-independent_2017}. However, in
these proof of principle experiments, the randomness estimation was
always evaluated in post processing after collecting all the raw data.
Here we implement a continuous variable SI-QRNG where all processing is
done in real-time. Additionally, we dynamically switch between two
measurement basis to alternate between a {\it check measurement} and a
{\it random-data measurement}. The SI-QRNG is self testing and changes its
output secure bit rate depending on the check measurement data. Although theoretical proposal for using squeezed states as sources of entropy for a QRNG
have been suggested~\cite{zhu_unbiased_2012, marangon_source-device-independent_2017}, we report the first
experimental use of squeezed states as an entropy source for a QRNG.

The paper is organized as follows. In section~\ref{sec:prot} we
present the protocol and experimental details for generating random
numbers. The protocol requires estimating a lower bound to the
conditional min-entropy. In section~\ref{sec:results}, we present the
real-time entropy estimation procedure and the statistics of the
random numbers generated. Due to finite sample size, we find that the
evaluated conditional min-entropy is positively biased which can lead
to an overestimation of the randomness rate. To mitigate this, we
propose and discuss other more robust estimators in
section~\ref{sec:estimators}. Finally, we conclude in
section~\ref{sec:conclusion} with a discussion of several ways for
extending the work in this paper as well as a summary of our work.

\section{Protocol and experiment}
\label{sec:prot}
\begin{figure*}[t!]
\centering
\includegraphics[scale=0.2]{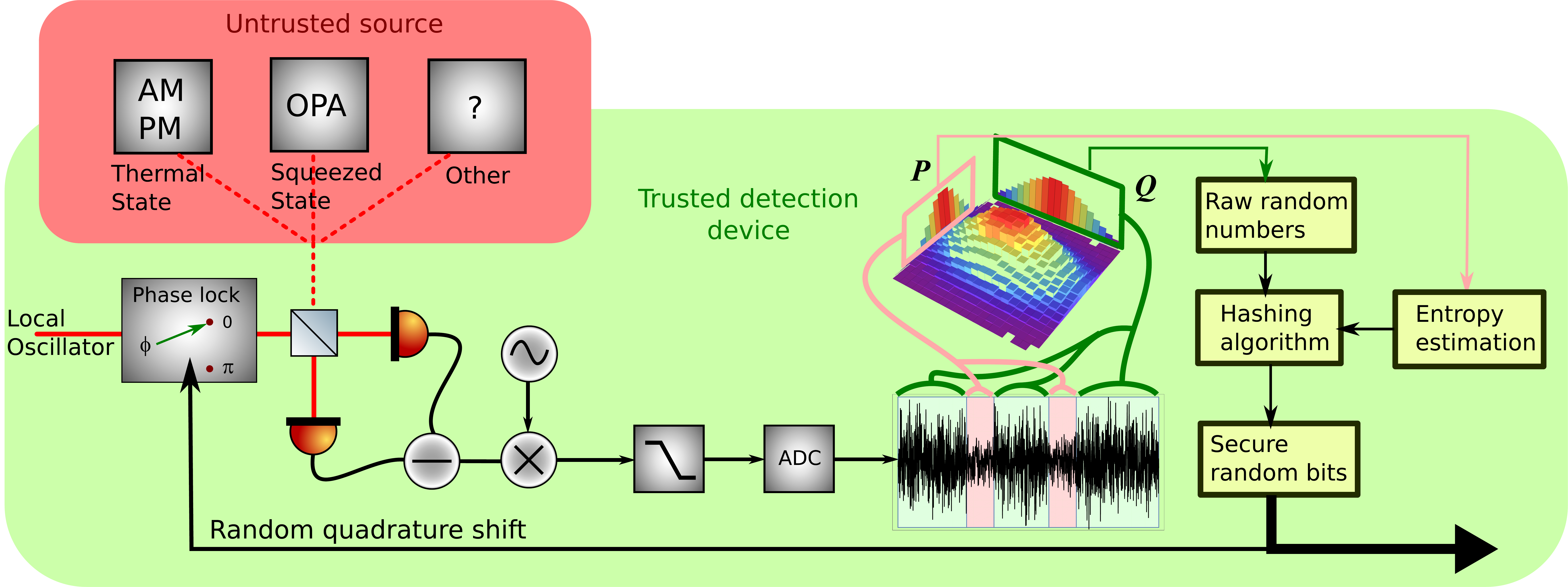}%
\caption{\label{setup}Scheme and protocol of the SI-QRNG. A 
  local oscillator whose phase is locked to measure the \textit{check}
  quadrature is interfered with an untrusted entropy source which can be a squeezed,
  thermal or some unknown state. The two output beams are detected and
  the resulting photocurrents are subtracted. From this homodyne measurement,
  the min-entropy on the \textit{random-data} quadrature is estimated. The phase lock then switches
  to the orthogonal random-data quadrature and the same homodyne
  measurement is performed. The raw random numbers are hashed according to the previous min-entropy
  estimation. Some of the secure random bits obtained in this way are
  used to determine when the next lock switch will happen. The check
  quadrature is measured randomly and on average once every ten runs.}
\end{figure*}

In a SI-QRNG, we are attempting to generate secure random numbers
without having to trust the source state. This is possible by
performing trusted measurements on two non-commuting observables. Our
experiment was performed on continuous variable light fields, and the
observables  measured are the field quadratures $\hat{Q}$ and $\hat{P}$. By measuring the
check quadrature $\hat{P}$, we put a bound on how much secure
randomness can be extracted on the orthogonal random-data
quadrature $\hat{Q}$. In the following, we provide the details on how a bound on
the random-data measurements can be calculated.

\subsection{Randomness bound from conditional min-entropy}
In our experiment, even though the quadrature observable has a
continuous degree of freedom, the data that is recorded is ultimately
discrete. The discretization size is determined by the finite
resolution of the digitizer. This finite resolution
implies that we do not measure the observables $\hat{Q}$ and $\hat{P}$,
but rather their discretized counterparts. Formally we measure a
subsection of the positive-operator valued measures (POVM) $\left\{ \hat{Q}_{\delta q}^{k} \right\}$
corresponding to
$k\in \left[ -\frac{m}{2}, \frac{m}{2}-1 \right]$, where
$\hat{Q}_{\delta q}^{k} = \int_{I_{\delta q}^{k}}{dq\ket{q}\bra{q}}$
and
\begin{align*}
  I_{\delta q}^{k} =\begin{cases}
\left( -\infty,
  (k+\frac{1}{2})\delta q\right)   & \text{for } k=-\frac{m}{2}\\
    \left[(k-\frac{1}{2})\delta q,
  (k+\frac{1}{2})\delta q\right)   & \text{for } -\frac{m}{2} < k <
\frac{m}{2}-1\\
    \left[(k-\frac{1}{2})\delta q,\infty \right)  & \text{for } k=\frac{m}{2}-1\;.
\end{cases}
\end{align*}
The even integer $m$ denotes the total number of bins, the index $k$
enumerates the outcomes and $\delta q>0$ specifies the precision of
the measurement. The measurement outcomes $q_{k}$ on state $\rho_A$, appear with
probability
$\mathfrak{p}(q_{k}) = \textrm{Tr}[\rho_{A}\hat{Q}_{\delta q}^{k}]$
and are stored in a classical register $Q_{\delta q}$.

When measuring the discretized  quadrature $\hat{Q}_{\delta q}$, the
maximum amount of secure
extractable randomness from a single shot measurement of
$Q_{\delta q}$ on a
state $\rho$ is given by~\cite{renner2008security,konig_operational_2009, koenig_sampling_2011,tomamichel_hierarchy_2013,frauchiger_true_2013,tomamichel_quantum_2016}
\begin{equation}
  \label{eq:ell}
  r_\text{sec}^{\epsilon}(Q_{\delta q}|E)_{\rho} = H_{\textrm{min}}(Q_{\delta q}|E)_{\rho} - 2\log_{2}\frac{1}{\epsilon}\;,
\end{equation}
where $\epsilon$ is the security parameter and
$H_{\textrm{min}}(Q_{\delta q}|E)$ is the conditional min-entropy of $Q_{\delta q}$~\cite{renner_security_2005}.  The protocol is then said to be $\epsilon$-secure, which means that the probability of distinguishing the output from a truly uniform independent distribution is smaller than
$\frac{1}{2}(1+\epsilon)$~\cite{tomamichel_quantum_2016}. The
conditional min-entropy $H_{\textrm{min}}(Q_{\delta q}|E)$ is defined
as~\cite{konig_operational_2009,tomamichel_fully_2009,koenig_sampling_2011}
\begin{equation}
 H_{\textrm{min}}(Q_{\delta q}|E)_{\rho_{QE}} = -\log_2 \max_{\{\hat E_k\}} \underbrace{\sum_k \mathfrak{p}(q_{k}){\rm Tr}[\hat E_k \rho^{k}_{E}]}_{p_\text{guess}(\{\hat{E}_k\})}\;,
 \label{eq:con_hmin}
\end{equation}
where
$\rho_{QE} = \sum_k \mathfrak{p}(q_{k})\ket{k}_A\bra{k}\otimes
\rho^{k}_E$ is the collective classical-quantum state after
measurement on system $A$. $ \rho^{k}_E$ is the state of system $E$
conditioned on measurement $k$ by $A$, and $\{\hat E_k\}$ is a POVM on
system $E$. The quantity $p_\text{guess}(\{\hat{E}_k\})$ is the average probability
for an adversary Eve to correctly guess the index $k$ using a measurement strategy
$\{\hat{E}_k \}$. The maximization on the POVM $\{\hat E_k\}$ corresponds to
finding the best measurement strategy Eve might apply to guess the
index $k$ on the post-measurement state $\rho_{QE}$. The amount
  of secure randomness is then the smallest conditional min-entropy
  for states $\rho_{QE}$ consistent with Alice's state $\rho_A$. If
the state $\rho_{A}$ is pure, this implies that $A$ and $E$ are independent:
$\rho_{AE}=\rho_{A}\otimes\rho_{E}$, in which case the conditional min-entropy
reduces to the classical unconditioned min-entropy
\begin{equation}
H_{\textrm{min}}(Q_{\delta q}) = -\log_{2} \max_{k}\{\mathfrak{p}(q_{k})\}\;.
\end{equation}
Here Eve's best guessing strategy is to guess the most likely index
$k$ everytime. For any state,
$H_{\textrm{min}}(Q_{\delta q}) \geq H_{\textrm{min}}(Q_{\delta q}|E)$
and the difference can be seen as the amount of side information
detained by Eve. To compute the exact value of
  $H_\textrm{min}(Q_{\delta q}|E)_{\rho_{QE}}$ in (\ref{eq:con_hmin}),
  one needs to know $\rho_{QE}$. Since Alice does not have access to
  $E$, she would need to perform a complete tomography of $\rho_A$ to
  find all compatible states $\rho_{QE}$. This is tedious for an
  infinite dimensional system. Instead, one can bound
$H_{\textrm{min}}(Q_{\delta q}|E)$ by the max-entropy of the conjugate
quadrature $H_{\textrm{max}}(P_{\delta p})$ using the \textit{entropic
  uncertainty relation} (EUR) \cite{bialynicki-birula_uncertainty_1975,berta_uncertainty_2010,rastegin_entropic_2011,tomamichel_uncertainty_2011,coles_uncertainty_2012,furrer_position-momentum_2014,zhang_renyi_2015,coles_entropic_2017}. This leads
to a lower bound $H_\text{low}$:
\begin{equation}
\label{eqn:hlowdef}
H_{\textrm{min}}(Q_{\delta q}|E) \geq H_{\textrm{low}}(P_{\delta p}) \coloneqq - H_{\textrm{max}}(P_{\delta p}) -\log_{2}c(\delta q,\delta p)\;
\end{equation}
where the max entropy is defined as
  \begin{equation}
    \label{eqn:hmaxdef}
H_{\textrm{max}}(P_{\delta p}) = 2\log_{2} \sum_{k}{\sqrt{\mathfrak{p}(p_{k})}}\;.
\end{equation}
The classical unconditioned max and min entropies are equivalent to the
R{\'e}nyi entropies \cite{Ren61} of order $\frac{1}{2}$ and $\infty$
respectively. Additionally,
\begin{equation}
c(\delta q, \delta p) = \frac{1}{4\pi}\delta q\delta p S^{(1)}_{0}\left(1,\frac{\delta q\delta p}{8}\right)^{2}
\end{equation}
is a measure of the incompatibility between the two measurements and
$S^{(1)}_{0}$ is the $0$th radial prolate spheroidal wave function of
the first kind \cite{landau_prolate_1961}. This wavefunction comes
about by considering the maximum overlap between the eigenstates of
$\hat{Q}_{\delta q}$ and $\hat{P}_{\delta p}$ and holds for states
$\rho_A$ that has no support in the two extreme bins because those extreme bins have width
larger than $\delta q$. This requirement correspond to bounding the
energy of the input state which is a reasonable assumption. This
assumption could be verified by including an energy measurement as part
of the protocol~\cite{furrer_reverse-reconciliation_2014, zhang_continuous-variable_2018}. In our real-time demonstration, we just limit the excursion of the state by aborting the protocol when one of the extreme bins is populated.

The bound is unconditional; it is a function of
$P_{\delta q}$ only and is independent of $E$. We use the convention where the vacuum state has a quadrature variance of
1. The $\hat{P}$ quadrature is used as the check quadrature that
estimates the amount of randomness extractable in the measurement on
the conjugate $\hat{Q}$ which is the random-data quadrature.

\subsection{Experimental details}

As shown in~Fig.\ref{setup}, the experimental set-up has two parts.
The fist part is an untrusted entropy source which consists of a
quantum state $\rho_{A}$ that may be mixed and correlated to a
malicious party E: $\rho_{A} = \text{Tr}_{E}(\rho_{AE})$.  We operated
the device on two sources, a squeezed state and a thermal state.
A shot-noise limited 1064 nm Nd:YAG continuous wave laser
provides the laser source for this experiment. A portion of the 1064
nm light is frequency doubled to provide a pump field at 532 nm. Both
fields undergo spatial and frequency filtering to provide shot-noise
limited light at the sideband frequencies above 2 MHz.  The thermal
state was generated with an amplitude and phase electro-optic modulators on
which we send a white noise electronic signal from two independent function
generators. By varing the amplitudes of the noise to the modulators, we varied
the variance of this thermal state to see the effect on the secure bit
rate. The squeezed state with around $3$ dB of squeezing was generated with a seeded doubly resonant
optical parametric amplifier in a bow-tie geometry. Details on the squeezed state generation can be found in~\cite{Chrzanowski_2013}.

The second part of the set-up is a trusted measurement device which consists of a homodyne detector that can measure one of two conjugate quadratures $\hat{Q}$ and $\hat{P}$ on the state $\rho_{A}$
by locking the phase of a local oscillator (LO) using amplitude or phase
modulation. The subtracted current is then mixed down and filtered
in the 13 to 17 MHz band before being digitized over $m=2^{12}$
bins. The measurement device switches randomly between two
measurement states: check measurements and random-data
measurements. On average, a check measurement was performed once every
ten measurement cycles.

In the check measurement state, three measurement steps are
performed. In the first step, the LO and signal beams are blocked using servo-controlled beam-blocks and
the electronic dark noise is recorded. In the second step, the signal
beam is blocked, while the LO is unblocked. This allows us
to record the vacuum shot noise. In the third step, both signal and LO
beams are unblocked; the LO is locked to $\hat{P}$ and the check data
is recorded. The data is then normalized according to the shot noise
corrected of dark noise $\sigma_{\text{corrected}}^2 = \frac{\sigma_{\text{measured}}^2}{\sigma_{\text{shot}}^2-\sigma_{\text{dark}}^2}$. In this way, all electronic noise will be accounted as impurity on $\rho_{A}$.

From the check data, we evaluated the probabilities
$\mathfrak{p}(p_k)$ using the frequentist estimator and
$H_{\textrm{max}}(P_{\delta p})$ using eqn~(\ref{eqn:hmaxdef}). For
each evaluation, the bin size $\delta p$ is recalculated, in units of
shot noise, using the corrected shot noise measurement. The
corresponding value of $c(\delta q,\delta p)$ is then evaluated using
a pre-calculated polynomial approximation. In the experiment, we had
an average $\delta q = (14.45 \pm 0.09) \times 10^{-3}$ and
$\delta q = (15.56 \pm 0.09) \times 10^{-3}$ for the thermal state and
squeezed state run respectively. The bound
$H_\textrm{low}(P_{\delta p})$ is then estimated
using~(\ref{eqn:hlowdef}) and stored in the computer for use in the
random-data measurement stage.

In the random-data measurement state, both the signal and LO beams are
unblocked. The LO phase is locked to $\hat{Q}$, and the raw data is
recorded. It is then normalized according to the shot noise corrected
of dark noise taken from the previous check measurement. In order to
eliminate Eve's information, we apply the Toeplitz matrix hashing
algorithm~\cite{ma_postprocessing_2013} on the raw data to obtain the
secure random data. The length of the Toeplitz matrix is determined by
the randomness bound evaluated in check stage. A few bits of the
hashed random numbers are used to determine whether the next stage
will be a check or random-data measurement stage.

For both check and random-data measurements, the number of
measurements points collected was $n = 16 000$. In our implementation,
to avoid slowing down the protocol, the random Toeplitz matrix was
generated once at the start of the experiment using a trusted 
QRNG~\cite{haw_maximization_2015} source. However, for the hashing to be
fully secure, a new hashing function randomly chosen from a family of
two-universal hashing functions should be used every time~\cite{bennett1995generalized, Renner2005,
  tomamichel_quantum_2016}. This is so that Eve does not have
knowledge of the hash function prior to preparing the state such
that she cannot implement deception strategies tailored to the hashing
function. For monitoring purposes, we also evaluated
$H_{\textrm{min}}(Q_{\delta q})$ using the frequentist estimator. We
also check that the homodyne detection was never saturated. If a
saturation event is ever detected, the protocol will abort
immediately.

\section{Results and Estimation error analysis}
\label{sec:results}

\begin{figure*}[t]
	\label{Hlow_exp_rt}
	\begin{subfigure}[b]{0.5\textwidth}
		\centering      \includegraphics[width=0.95\textwidth]{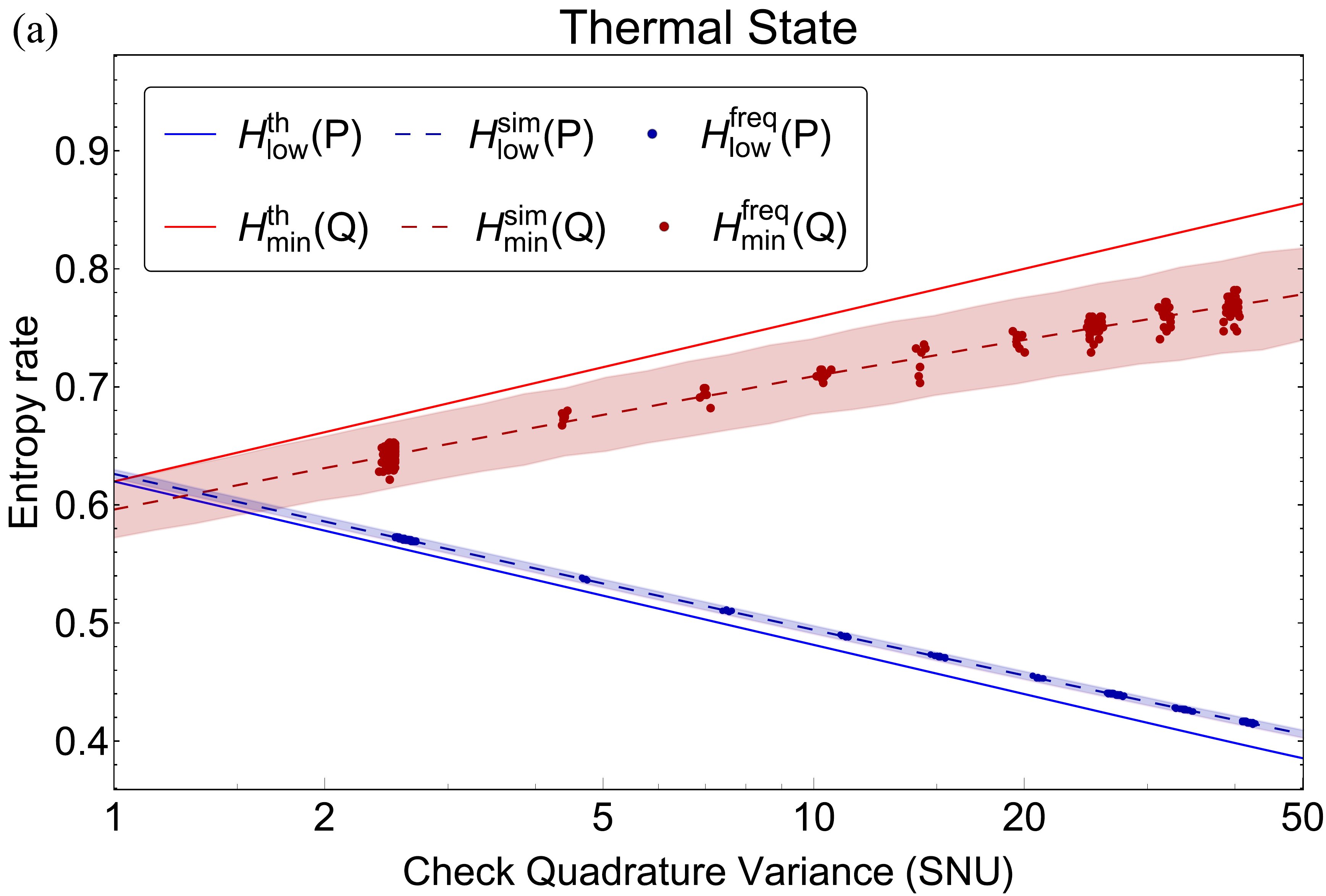}%
		\caption{\label{Hlow_exp_th}}
	\end{subfigure}%
	~ 
	\begin{subfigure}[b]{0.5\textwidth}
		\centering
		\includegraphics[width=0.97\textwidth]{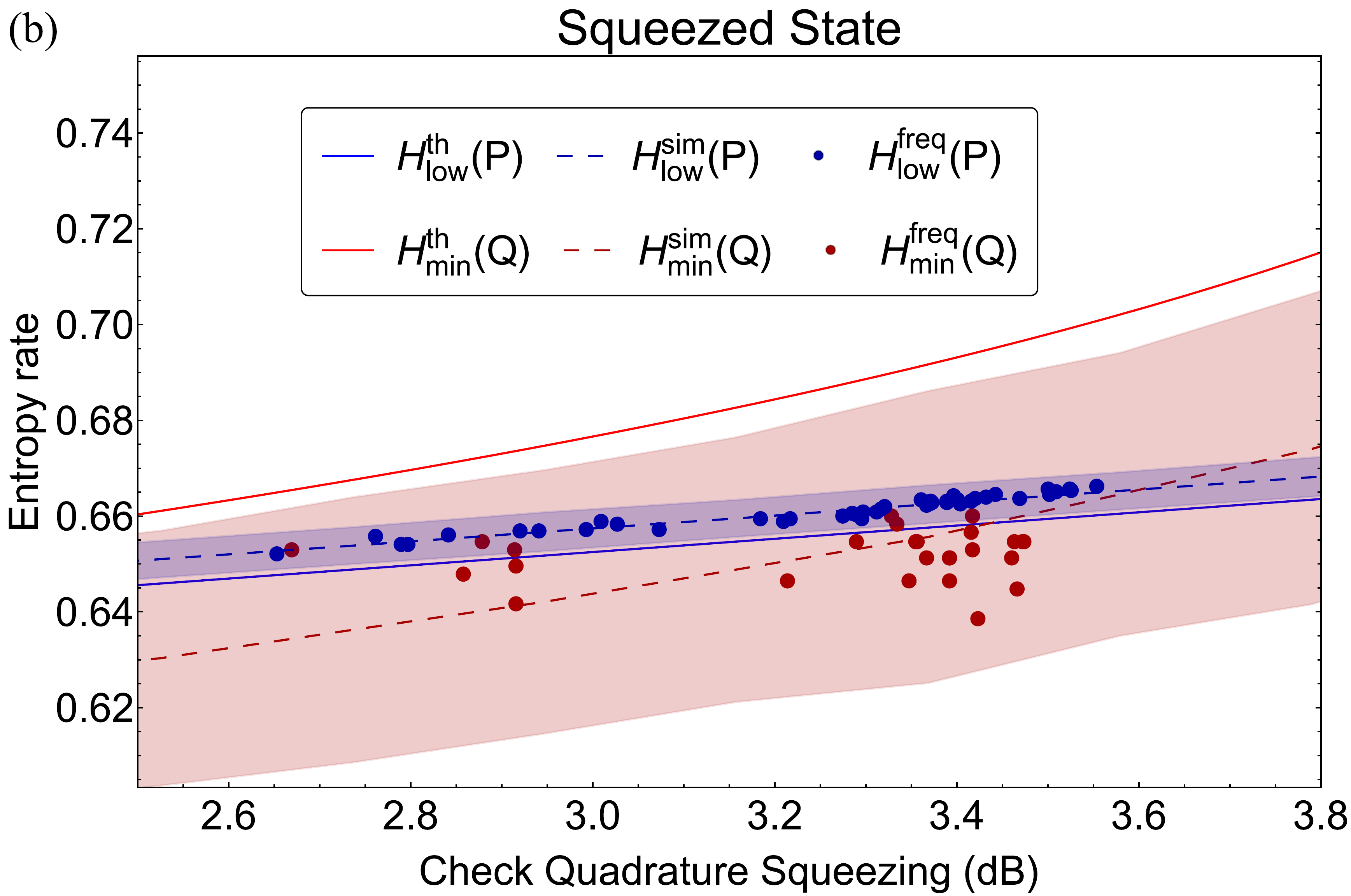}%
		\caption{\label{Hlow_exp_sqz}}
	\end{subfigure}
	\caption[]{ Entropy bound and classical min-entropy for (a) a
          thermal state with different values of noise and (b) a
          $\hat{P}$-squeezed state with $33\%$ loss. The red solid lines
          show the theoretical unconditioned min entropy of the random-data
          quadrature $\hat{Q}$. This gives the extractable randomness if the
          source is trusted. The blue solid lines show the theoretical
          bound to the conditional min-entropy $H_\textrm{min}(Q|E)$
          obtained by the entropic uncertainty relation. This gives
          the secure extractable randomness for an untrusted source.
          The blue and red points show the corresponding experimental
          data calculated in real-time using a frequentist estimator
          on data samples of length $n=16000$. For most values of
          squeezing, we find that
          $H^\text{freq}_\textrm{low}>H^\text{freq}_\textrm{min}$
          which would be in violation of the EUR (eqn~\ref{eqn:hlowdef}). This apparent
          violation arises due to a bias in the frequentist
          estimators. Dashed lines show the corresponding simulation
          results and the shaded area corresponds to a $5$ standard
          deviation uncertainty region.}
\end{figure*}

As mentioned before, the QRNG was operated with two different sources;
a $\hat{P}$-squeezed state and a thermal state. In order to generate
secure randomness, we use the bound provided by
$H_\text{low}(P_{\delta p})$ in eq~(\ref{eqn:hlowdef}). In order to
apply this bound, we need to know the value of
$H_\text{max}(P_{\delta p})$. In subsection~\ref{Real time entropy
  estimation}, we present the real-time experimental result where the
frequentist estimator for $H_\textrm{max}(P_{\delta p})$ was used. In
subsection~\ref{Bias of the frequentist estimator}, we show that this
estimator is biased which may compromise the security of the QRNG.

\subsection{Real time entropy estimation}
\label{Real time entropy estimation}

In the experiment, the entropies are calculated in real time using the
frequentists estimator. After measuring $n=16000$ data points, and
binning the outcomes into $m$ bins, the frequentist estimators are given by
\begin{align}
{H}_{\textrm{min}}^{\textrm{freq}}(\vec{n}) &=
                                                      -\log_{2}\frac{\max_{k}\{n_{k}\}}{n}\;,\\
{H}_{\textrm{max}}^{\textrm{freq}}(\vec{n}) &= 2\log_{2}\sum_{k=1}^{m}{\sqrt{\frac{n_{k}}{n}}}
\end{align}
where $n_k$ denotes the number of outcomes in the $k$-th bin and
$\vec{n}=(n_1,n_2,\ldots,n_m)$.  The entropy bounds
$H^\text{freq}_\textrm{low}(P_{\delta p})$ and the unconditioned
classical entropy $H^\text{freq}_{\textrm{min}}(Q_{\delta q})$ from
the experiments are recorded for thermal and squeezed state and these
are presented as the points in Figs.~\ref{Hlow_exp_th} and
\ref{Hlow_exp_sqz}. In the same figures, we also plot simulation
results $H^\text{sim}_{\textrm{low}}(P_{\delta q})$ and
$H^\text{sim}_{\textrm{min}}(Q_{\delta q})$ obtained by sampling $n$
points from a perfect Gaussian distribution. These simulations are
repeated $1000$ times to estimate the mean and standard deviation of
the estimated entropy bound. Finally, the theoretical values we would
expect for a perfect discretized Gaussian distribution
\begin{align}
\mathfrak{p}(p_{k}) = \frac{1}{2}\textrm{erf}\left(\frac{p_{k}+\frac{\delta p}{2}}{\sqrt{2}\sigma}\right)-\frac{1}{2}\textrm{erf}\left(\frac{p_{k}-\frac{\delta p}{2}}{\sqrt{2}\sigma}\right)\;
\end{align}
are plotted as solid lines $H^\text{th}_{\textrm{low}}(P_{\delta q})$
and $H^\text{th}_{\textrm{min}}(Q_{\delta q})$.

For the squeezed state simulation results, the impurity of the
squeezed state was accounted for by inferring the amount of
loss on the state from the two quadrature variance measurement. This
was estimated to be $33\%$. This is the reason why the min-entropy and
the bound are not equal; they can only be equal for a pure
state. Figure \ref{Hlow_exp_sqz} shows that higher squeezing give rise
to more extractable randomness. Indeed, measuring squeezing on one
quadrature guarantees increased noise on the conjugate anti-squeezed
quadrature. Unlike in the thermal noise case, this noise is not
correlated to another system. For example having $5$ dB squeezing on
the source increases the entropy rate by around $10$\% compared to
vacuum. Therefore using a squeezed state as an entropy source can
improve the QRNG bit rate, especially with broadband squeezing. For
the squeezed state run the bit rate was $8.2$ kb/s.

The thermal state results in Fig.~\ref{Hlow_exp_th} illustrate the difference between the
conditioned and unconditioned min-entropy. Indeed a thermal state can
be purified by a two-mode squeezed state such that the
outcome of a measurement on that state may well be correlated to a
mode detained by Eve. This amount of quantum or classical side
information is the difference between the min-entropy, which quantifies
the entropy of the measurement distribution, and the conditional
min-entropy which quantifies the entropy given any possible side
information. For a thermal state, the higher the variance the
higher the min-entropy which reflects the apparent random noise in quadrature
measurement, yet the lower the conditioned min-entropy because the
state could be a two-mode squeezed state with higher correlations. For the thermal state run, the bit rate varied between $5.2$
kb/s for the state with higher variance to $7.2$ kb/s for the state
with lower variance.

\begin{figure}[h!]
	\centering
	\includegraphics[width=0.475\textwidth]{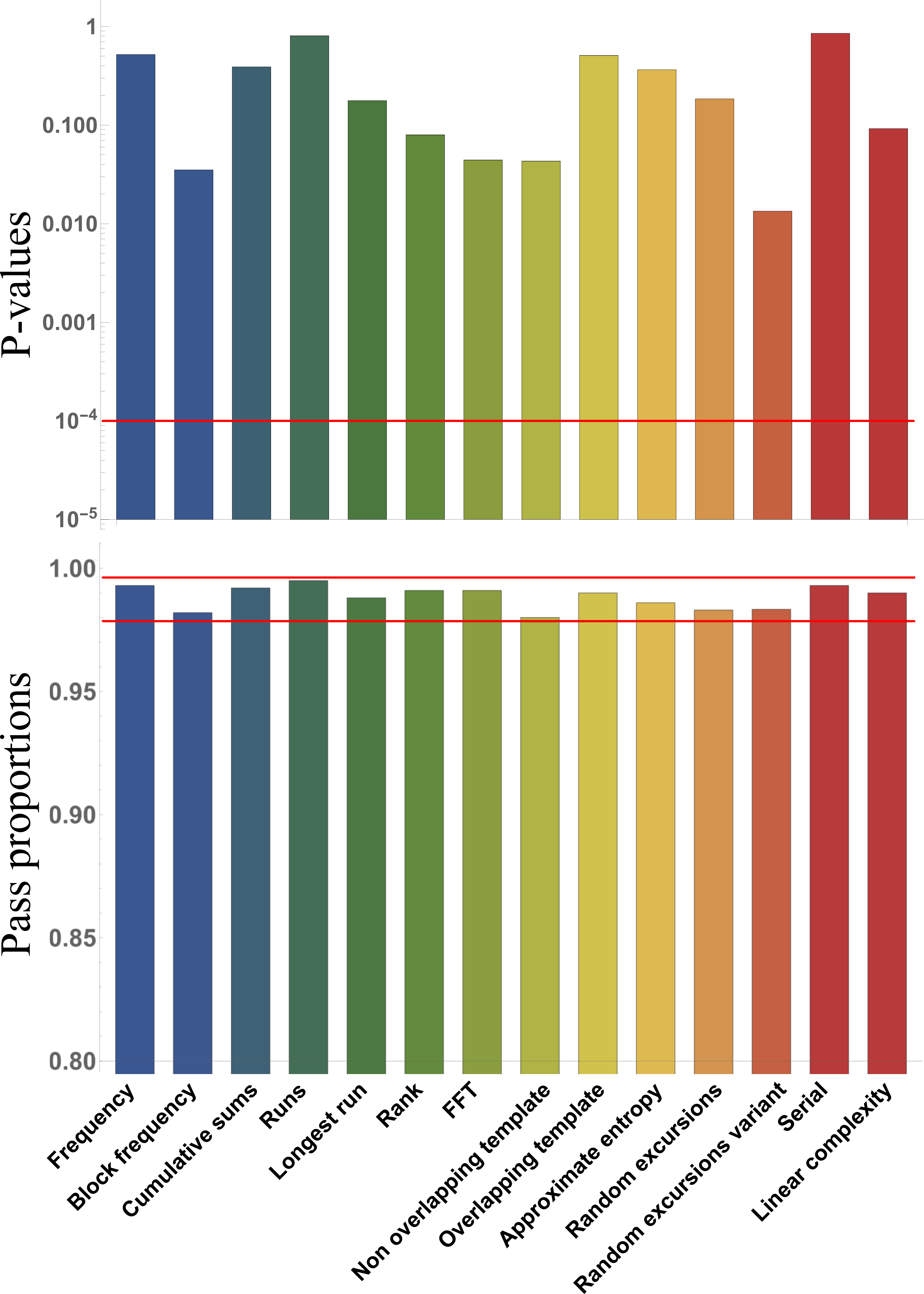}%
	\caption{\label{plot_NIST}Results of NIST statistical test
          suites obtained by combining $1000$ samples from both the
          squeezed and thermal state source. Each sample size is $100$ kbits
          and the test significance level is $\alpha=0.01$. To `pass', the
          $P$-values (uniformity of $p$-values) should be larger than
          $0.0001$ and the pass proportions should be within the
          Clopper--Pearson interval of
          $\left[0.978724, 0.996273\right]$.}
\end{figure}

We tested a collection of random numbers obtained from both the
thermal and squeezed states with the NIST statistical test suite
\cite{rukhin2001statistical}. The results are shown in
Fig.~\ref{plot_NIST}.

\subsection{Bias of the frequentist estimator}
\label{Bias of the frequentist estimator}
\begin{figure*}[t]
	\centering
	\begin{subfigure}[b]{0.5\textwidth}
		\centering
		\includegraphics[width=0.95\textwidth]{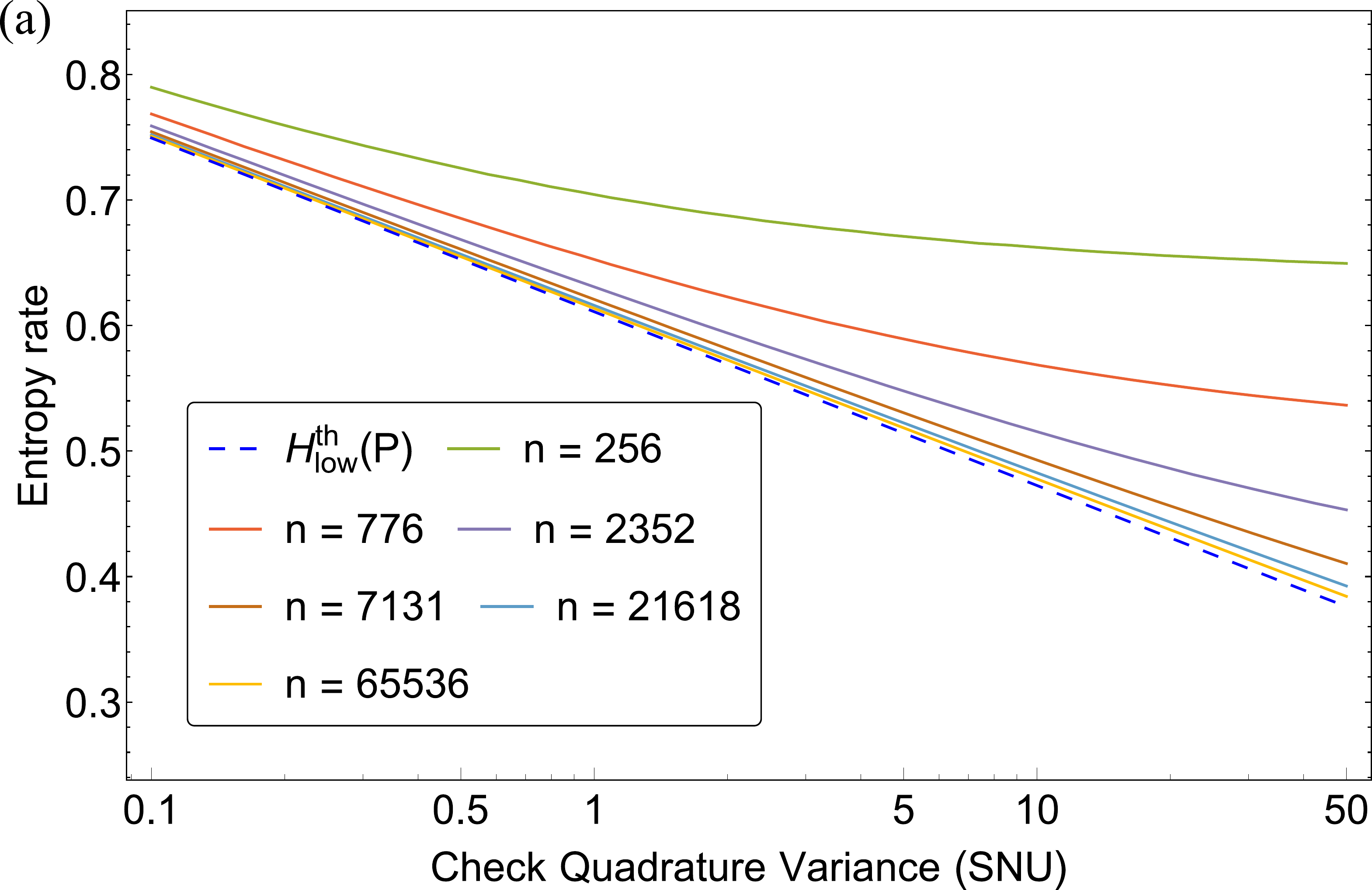}%
		\caption{\label{Hlow_sim}}
	\end{subfigure}%
	~ 
	\begin{subfigure}[b]{0.5\textwidth}
		\centering
		\includegraphics[width=0.95\textwidth]{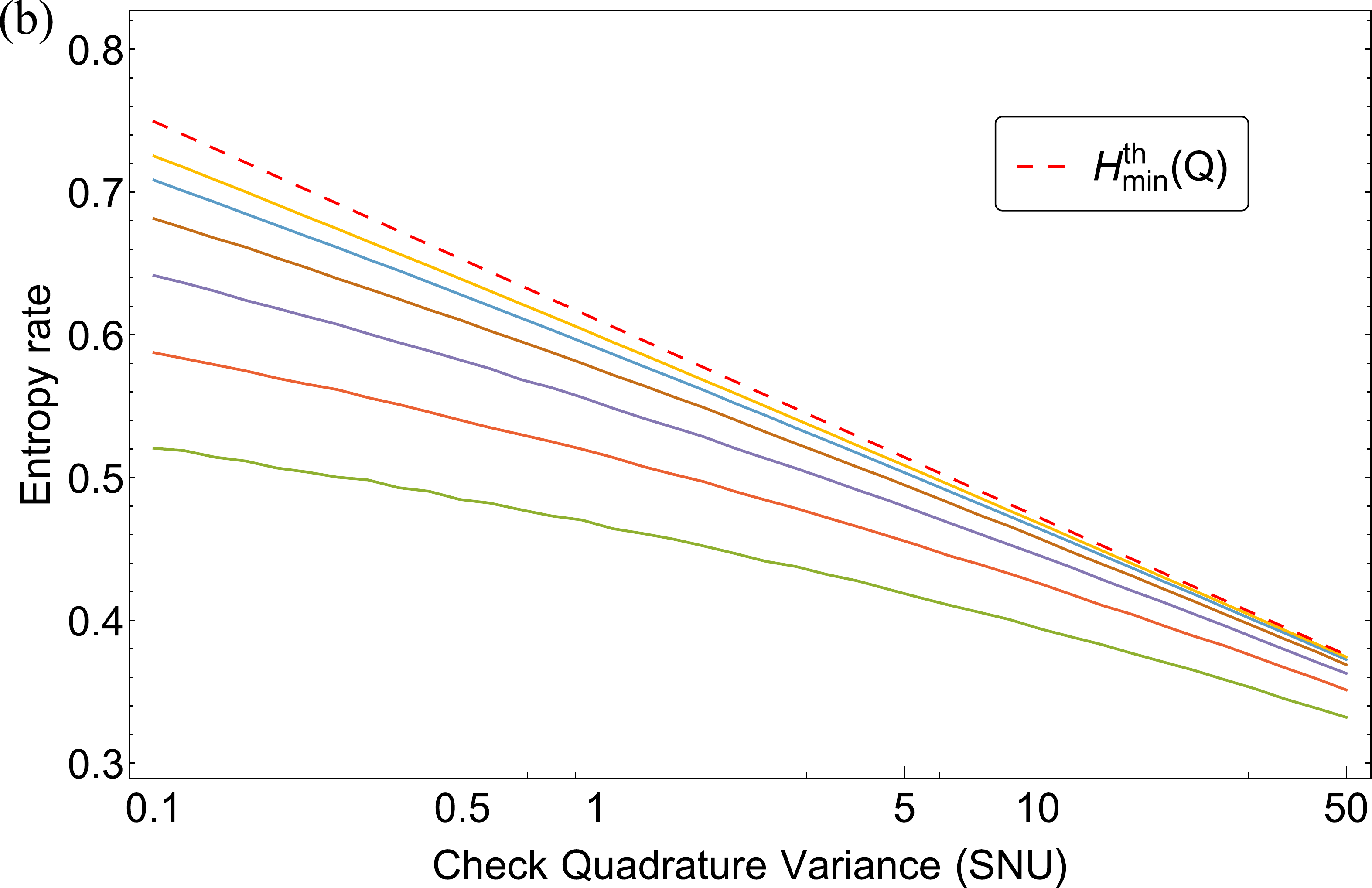}%
		\caption{\label{Hmin_sim}}
	\end{subfigure}
	\caption{\label{fig:freq_sim} (a) Simulation of the frequentist estimator of
          the entropy bound for a pure gaussian
          state. We set $\delta q=0.0155607$ which was the mean value
          of $\delta q$ for the squeezed state runs, and ran the simulation
          for different sample size. The dashed line shows the
          theoretical value of $H_\textrm{low}$ which gives a lower
          bound on the conditional min entropy. Due to the finite
          sample size, the estimator is positively biased which may
          lead to erroneously extracting more keys than is secure. (b)
          Simulation of the frequentist estimator of the unconditional min entropy with the
          same parameters. Due to finite sample size, this estimator
          is negatively biased which lead to instances where
          ${H}^\text{freq}_\textrm{min}<{H}^\text{freq}_\textrm{low}$.  }
\end{figure*}

We see in Figs. \ref{Hlow_exp_th} and \ref{Hlow_exp_sqz} that there is
a discrepancy between the theoretical bound
$H^{\textrm{th}}_{\textrm{low}}(P_{\delta p})$ and
$H^{\textrm{th}}_{\textrm{min}}(Q_{\delta q})$ calculated for a
Gaussian state and the experimental data. To analyse this we ran a
simulation by sampling a pure Gaussian distribution for different
sample size $n$. Each simulation was repeated $1000$ times. As shown
in Figs.~\ref{Hlow_sim} and \ref{Hmin_sim}, we find that the
frequentist estimators
$H^{\textrm{freq}}_{\textrm{low}}(P_{\delta p})$ and
$H^{\textrm{freq}}_{\textrm{min}}(Q_{\delta q})$ are both biased. The
mean of the frequentist estimators do not match the true values
$H_\textrm{low}^{\textrm{th}}(P_{\delta p})$ and
$H_\textrm{min}^{\textrm{th}}(Q_{\delta q})$. This leads to an
apparent violation of the EUR as
$H^{\textrm{freq}}_{\textrm{low}}(P_{\delta p})$ is positively biased
while $H^{\textrm{freq}}_{\textrm{min}}(Q_{\delta q})$ is negatively biased. This
bias gets smaller as the sample size increases. But even for very
large sample size this problem might be present, as it will depend on the
source state considered, as shown in Appendix~\ref{sec:smalln}.
Moreover if Eve's state is maximally correlated with ours then any overestimation of the bound will compromise the
security of the random numbers. One may try to correct this by using a
different estimator for the max-entropy.

\section{Other estimators for the entropy bound}
\label{sec:estimators}
Having learned that the frequentist estimator can be biased, in this
section we investigate and compare three different estimators. These
estimators comes with their own natural confidence interval that we
can set.

\subsection{Bayesian estimator}
Another class of possible estimators for
$H_{\textrm{max}}$ are the Bayesian estimators. To
calculate the Bayes estimator of an unknown parameter, one has to
specify a prior probability density. This represents our initial
belief about the distribution of the unknown parameter. Here we analyse two estimators
for $H_\text{max}$ based on two different priors. The first is an
uninformative prior which makes no assumption on the underlying
probability distribution. The second assumes the worst case
scenario by choosing a prior peaked around the
uniform probability. Deciding which prior to use is a matter of the
experimentalist's degree of paranoia. Even though the QRNG device is
source-independent, it is not belief-independent. We note that using Bayesian estimators bring with it the
additional advantage of having the posterior estimate as a natural confidence interval.

\subsubsection{Bayesian estimator for $H_\textup{max}$ with a completely uninformative prior}
\begin{figure}[t!]
	\centering
	\includegraphics[width=0.475\textwidth]{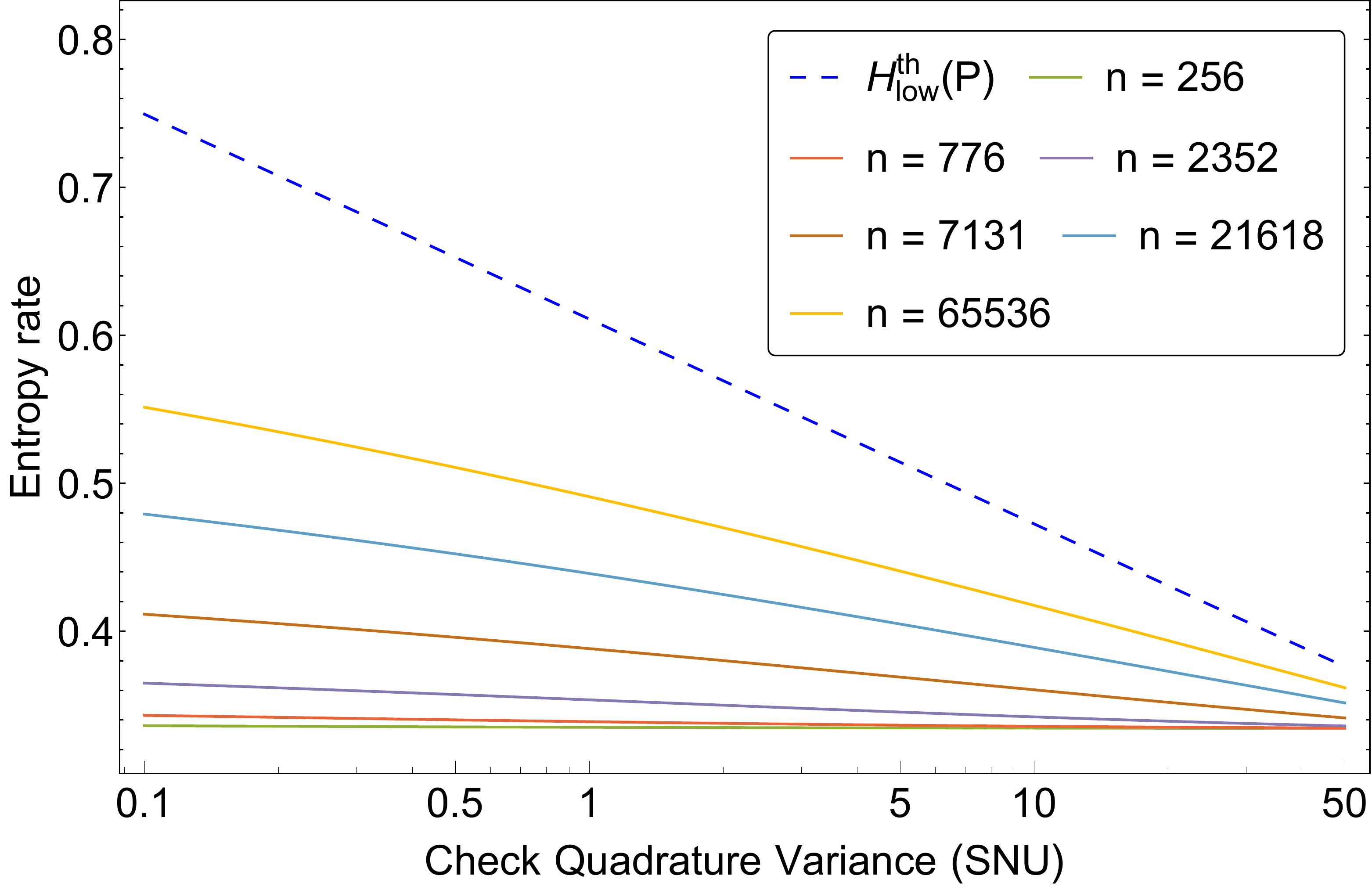}%
	\caption{\label{plot_hlow_bayes} Simulation of the uniform
          prior Bayesian estimator for the entropy bound of a pure
          gaussian state with the same parameters as
          Fig.~\ref{fig:freq_sim}. The estimator is negatively biased
          which does not compromise security. }
\end{figure}

The indirect Bayesian estimator with a completely uninformative uniform prior was developed in~\cite{wolpert_estimating_1995, holste_bayes_1998} and proposed
for source-device-independent QRNG in \cite{vallone_quantum_2014}. It is given by:
\begin{equation}
{H}^\text{up}_{\textrm{max}}(\vec{n}) =
2\log_{2}\left(\frac{\Gamma(n+m)}{\Gamma(n+m+\frac{1}{2})}\sum_{k=1}^{m}{\frac{\Gamma(n_{k}+\frac{3}{2})}{\Gamma(n_{k}+1)}}
\right)\;.
\end{equation}
Using this estimator in simulation for a gaussian state in our experimental conditions, we find that it has a negative bias which does not lead to a violation of the EUP (see Fig.~\ref{plot_hlow_bayes}). 
If one can check that the distribution is Gaussian, it is then
justifiable to use the Bayesian estimator. In fact, one can go a step
further and remove this bias from the estimator. Otherwise, this negative
bias will lead to a severe underestimation of the secure bit
rate. But a priori the distribution might not be Gaussian and this
bias will depend on the distribution and experimental conditions such
as the binning size. We show in Appendix~(\ref{sec:smalln}) that in
some extreme cases this bias can still be positive.

\subsubsection{Bayesian estimator for $H_\textup{max}$ with a prior
  peaked around the uniform distribution}
The Bayesian estimator depends on the chosen prior. The natural choice
of prior is the Dirichlet distribution since it is the
conjugate prior to the multinomial distribution. The Dirichlet
distribution parameterised by the vector $\vec\alpha$ is given by
\[\text{Dir}\left[\vec{\mathfrak{p}};{\vec{\alpha}}\right] =
\frac{\Gamma \left(\sum_{j=1}^m \alpha_j \right)}{\prod_{j=1}^m
  \Gamma(\alpha_j)} \prod_{j=1}^m \mathfrak{p}_j^{\alpha_j-1} \;,\]
where $\mathfrak{p}_j=\mathfrak{p}(p_j)$. In order to
prevent an under-estimation of $H_\text{max}$, it is prudent to assume
the worst case scenario by choosing a prior that is sharply peaked around
the uniform distribution. This is because the uniform distribution is
the distribution with the maximum possible $H_\text{max}$. We
subsequently adjust our belief when presented with the measured
data. Such a prior can be constructed by choosing $\alpha_j=K$ for all $j$
\begin{align}
  \pi({\vec{\mathfrak{p}}})&= \text{Dir}\left[ \vec{\mathfrak{p}};
                             K \right]\\
  &= \frac{\Gamma(mK)}{\Gamma(K)^m}(\mathfrak{p}_1\cdots
	\mathfrak{p}_m)^{K-1}\,.
	\label{prior}
\end{align}
Here $K$ characterize the peakedness of the prior distribution. A
large value of $K$ will correspond to a distribution peaked
around the uniform distribution, whilst $K=0$ will correspond to the
frequentist estimator. The Bayes posterior estimator given the
measurement outcomes $\vec{n}$ is the Dirichlet distribution with
parameters $\vec{\alpha}=\vec{n}+K$~\cite{Leonard1977}
\begin{align}
f(\vec{\mathfrak{p}}|\vec{n}) = \text{Dir}\left[ \vec{\mathfrak{p}}; \vec{n}+K\right]\,. 
\end{align}
From this posterior distribution, we can arrive at a Bayesian
estimator for $H_\text{max}$. Alternatively, an indirect estimator for
$H_\text{max}$ which we denote by $H^\text{pp}_\text{max}$ can be
obtained by substituting the Bayes posterior mean for the probabilities
$\vec{\mathfrak{p}}$
\begin{align}
  \label{ppp_k}
  \mathfrak{p}_j^\text{pp} &= \mathbb{E}\left[\mathfrak{p}_j|\vec{n}\right]\\
                        &= \frac{n_j+K}{n+m\,K}
\end{align}
into~(\ref{eqn:hmaxdef}). As we shall see in
section~\ref{subsec:estimator_comparison}, with a large $K$, this estimator tends
to be very conservative.

\subsection{Extremal Variance-based Estimator}
Another way to estimate $H_\text{max}$ is by estimating the variance
distribution. Instead of estimating $H_{\textrm{max}}(P_{\delta p})$
from the sampled distribution, we can try to bound it. We first
estimate $V_{P}$, the variance of $P_{\delta p}$ with the unbiased
estimator $V_P = \frac{1}{n-1}\sum_{k=1}^{n}(p_k-\bar{p})^2$. We can
then find the distribution that maximizes $H_{\textrm{max}}$ for this
given variance. This is similar to the method used in
\cite{xu_high_2017} for bounding the Shannon entropy
\cite{shannon_mathematical_nodate,wolf_extremality_2006}.

\begin{figure}[t]
	\centering
	\includegraphics[width=0.475\textwidth]{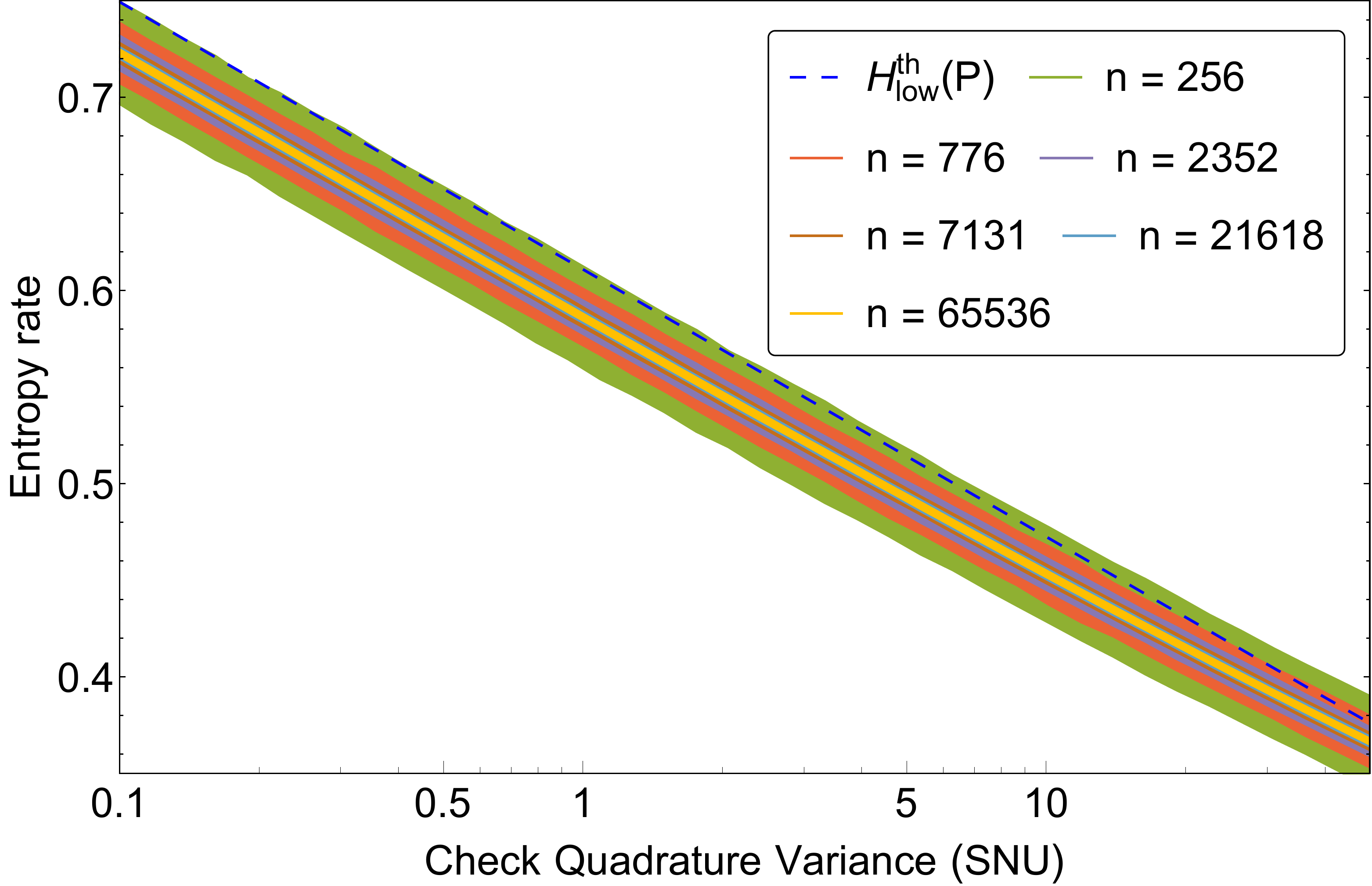}
	\caption{\label{Hlow_sim_extr}The extremal variance based
          estimator $H_\textrm{low}^\textrm{EVB}$ obtained by
          estimating the variance of the check-quadrature. Shaded area shows $5$ standard deviation. This
          estimator shows no bias. The dashed line is the
          theoretical bound for a Gaussian
          distribution. The estimator is lower because the
          extremal distribution for the EVB estimator assumes a
          discretised Student's t-distribution (eqn~\ref{eq:dst}).  For
          sample size above $1000$, the variance of this
          estimator become small enough so that the probability of a
          single shot estimation to be above
          $H_\textrm{low}^\textrm{th}$ becomes negligible.}
\end{figure}

We show in Appendix~\ref{Appendix_extremal_proof} that given a
variance $V_p$, the corresponding extremal
distribution is given by
\begin{equation}
  \label{eq:dst}
 \mathfrak{p}(p_{k}) = C \frac{1}{\left(1+\left(\frac{p_k}{s}\right)^2\right)^{2}}
\end{equation}
where
\begin{align}
  C = \sum_j \frac{1}{\left(1+\left(\frac{p_k}{s}\right)^2\right)^{2}}
\end{align}
is a normalization constant,
\begin{align}
 s = \sqrt{\frac{1-\gamma V_P}{\gamma}} 
\end{align}
 and $\gamma$ is the solution to the equation
\begin{equation}
\label{eg:gamma}
\sum_{k}\frac{p_k^2-V_P}{\left(1+\gamma (p_k^2-V_P) \right)^{2}} = 0\,.
\end{equation}
This distribution is a discretized Student's t-distribution with 3 degrees of freedom.
Although equation (\ref{eg:gamma}) does not have a closed form
solution for $\gamma$, one may
calculate it numerically. We can then calculate the extremal variance based (EVB)
estimator $H^{\textrm{EVB}}_{\textrm{max}}(V_{P})$. This is the extremal max-entropy
consistent with the variance $V_{P}$. From this we get an estimate for $H^\textrm{EVB}_\textrm{low}$ from
eqn~(\ref{eqn:hlowdef}). This is plotted in Fig.~\ref{Hlow_sim_extr} for a gaussian state
with parameters similar to our experiment.

In these conditions, we see that the EVB estimator shows no bias, the
mean value does not change with the sample size. Moreover, by
construction, the mean of the EVB estimator for
$H^\textrm{EVB}_\textrm{low}$ is always smaller than
$H_\textrm{low}^\textrm{th}(P_{\delta p})$.  Unlike the frequentist
estimators, the EVB estimator does not over-estimate
$H_\textrm{low}^\textrm{th}$. However, because the EVB estimator uses
only the variance instead of the whole distribution it does not
converge to $H_\textrm{low}^\textrm{th}$ even when the sample size is
large. It will only converge to $H_\textrm{low}^\textrm{th}$ if the
check-quadrature distribution happens to be the discretized Student's t-distribution~(\ref{eq:dst}).

We note that here, the theory
$H_{\textrm{low}}^{\textrm{th}}(P_{\delta p})$ and simulations were
computed for Gaussian states. The bias results will differ for other
input state and in some cases the EVB estimator can still be
positively biased. This is because even though the variance estimator
is unbiased, the max-entropy is a concave function of the
variance. This means that it will have a negative bias. This is
illustrated in Appendix~\ref{sec:smalln}. However, we can get a confidence interval on the
variance from the sampled data and from this we can arrive at a confident estimate for the
max-entropy.

\subsection{Comparison of the different estimator performances}
\label{subsec:estimator_comparison}
\begin{figure}[t]
	\centering
	\includegraphics[width=0.48\textwidth]{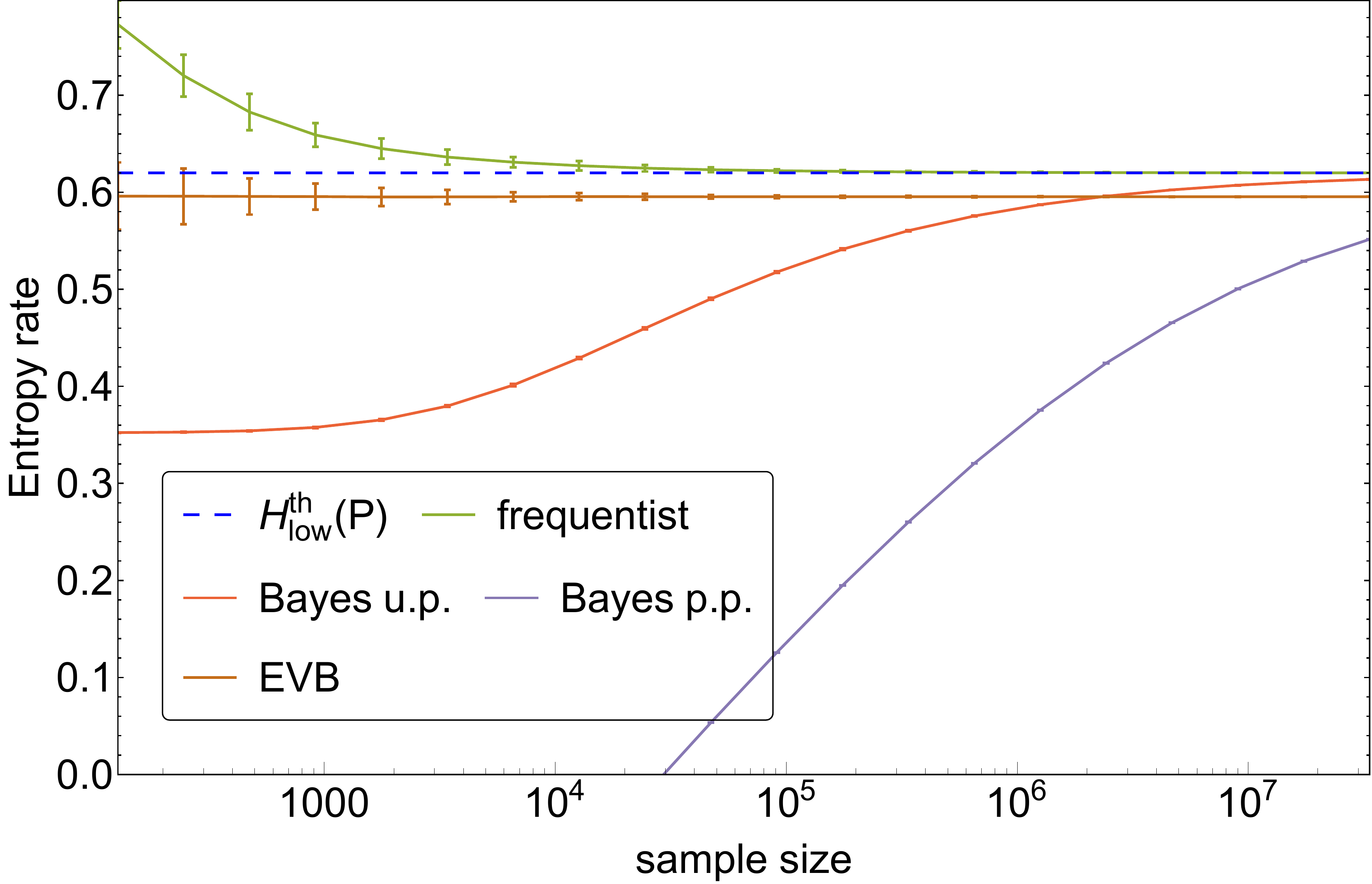}%
	\caption{\label{comparison} Comparison between frequentist,
          Bayes and EVB	estimators for $H_\text{low}$ with finite statistics for the
        vacuum state. For the Bayesian estimator with peaked prior, $K$ was set to
$100$. Each simulation was repeated $100$ times to obtain
        the estimator mean and standard deviation.}
\end{figure}

A comparison on how the different estimators perform with increasing
sample size for a vacuum state input is shown in
Fig.~\ref{comparison}. The frequentist estimator has a positive bias
leading to an overestimation of the secure randomness rate which can
compromise the security of the random numbers. In contrast, the EVB and both Bayes
estimators have a negative bias which leads to an underestimation of
the secure randomness rate. Of all the estimators, the Bayesian peaked
prior estimator is the most conservative, it will significantly
underestimate the bound even for large sample sizes.

Finally we note that even with an unbiased estimator for $H_\text{max}$,
one should not take its mean value as the point estimate. Doing this
will lead to a 50\% probability of over estimating $H_\text{max}$. Instead,
one should get a point estimate based on its confidence interval and a
required failure rate.

\section{Conclusion and outlook}
\label{sec:conclusion}
We demonstrated a real time source independent QRNG incorporating
measurement basis switching and hashing using a squeezed state of
light as a source of entropy. The protocol was also validated on
different thermal states. In the real time demonstration, the sample
size was limited by the finite computational resource. A valuable
lesson learnt from this demonstration is that due to finite size
effects, the frequentist estimator can lead to an underestimation of
the max-entropy due to its biased nature. This can lead to an
underestimation of the adversary's knowledge on the measured data. To
mitigate this potential problem, we propose three different ways for
estimation of the max-entropy. Which one of these estimators the
experimenter picks will depend on his level of paranoia.

We note that this estimation problem does not arise in a trusted source QRNG where confidence interval on entropy estimator can be calculated from the knowledge of the source. Nor does it appear in asymptotic CV quantum key distribution (QKD) protocols where the measured distribution can be assumed to be Gaussian due to optimality
of Gaussian attacks~\cite{garcia-patron_unconditional_2006,
  navascues_optimality_2006}. For Gaussian distributions, it is then
easy to construct a confidence interval for the max-entropy. However,
in a source independent protocol, we see that a Gaussian distribution
is not the best that the adversary can do. Hence assuming a Gaussian
distribution might lead to an underestimation of her knowledge.

The bit rate was limited by three main factors. First, the slow real-time
hashing of the raw bits which was done on a desktop computer. Second,
the mechanical beam blocking in check measurement. Third, the limited
squeezing bandwidth. The first
limitation can be circumvented using a fast programmable gate arrays
(FPGA) to hash. We foresee that implementing the hashing on an FPGA
would allow us to reach the GHz regime~\cite{zhang_note:_2016}. The second limitation is less
stringent since the beam blocking only happens during the
check-measurement. In our setup, the check-measurement was performed
with a $10\%$ probability and the data-measurement is not limited by
these slow mechanical beam blocks. Furthermore, one may use faster,
non mechanical ways to block the beam, for example by using acoustic
optical modulators to deflect the beams. The third limitation in this
experiment was the squeezing bandwidth which is imposed by the
bandwidth of the OPA squeezing cavity. Hence, using a squeezed state
source may limit the bit rate through bandwidth limitation more than
it improves it through the higher security rate. This limitation can
be circumvented by using a single pass OPA with would offer squeezing over much
larger bandwidths~\cite{ast_high-bandwidth_2013}.

In conclusion, we demonstrated a real-time source independent maximum 
bit rate of 8.2 kbits per second with a squeezed state source, and 
from 5.2 to 7.2 kbits per second with thermal source depending on the
variance of the source.

\section{Acknowledgement}
This work was funded by the Australian Research Council Centre of Excellence and Laureate Fellowship schemes (CE110001027 and FL150100019). Our research is also supported by The Defence Industry and Innovation Next Generation Technologies Fund.

We thank Nathan Walk for useful discussions and comments on this work.
\bibliographystyle{apsrev4-1}

\bibliography{SDIQRNG}

\pagebreak
\appendix

\section{Extremal distribution for max-entropy with a fixed variance}
\label{Appendix_extremal_proof}
Suppose we experimentally observed a discrete distribution in a finite
  support. From the variance of this distribution, we can upper
  bound its entropy. To do that we shall derive the probability distribution
that maximizes the entropy for a fixed variance. We note that
entropy does not depend on the labels of the bins; to have a tighter
bound we can rearrange the bins to minimize the variance.

Here we derive the probability distribution maximizing max-entropy for
a fixed variance in a finite support setting. We want to find the
extremal distribution $\mathcal{P}=\{p_k\}$ that maximizes the max
entropy:
\begin{equation}
H_{\textrm{max}}(\vec{p}) = 2\log_{2} \sum_k{\sqrt{p_{k}}}
\end{equation}
over the finite support $x_k = k\, \delta x$ for integer values
$k\in[-m,m]$ subject to the normalization constraint $\sum_k p_k=1$ and fixed variance condition:
\begin{equation}
    \sum_k p_k\, x_k^2 - \left(\sum_k p_k\, x_k  \right)^2=V\,.
\end{equation}
We first show that the extremal distribution must be symmetric with $p_k=p_{-k}$. From an arbitrary distribution $\mathcal{Q}=\left\{q_k\right\}$, we can construct a symmetrized distribution $\mathcal{P}=\{p_k\}$ with 
\begin{align*}
 p_{k} = \frac{q_{k}+q_{-k}}{2}\,.
\end{align*}
This distribution will have a smaller variance, $\text{var}(\mathcal{P}) \leq \text{var}(\mathcal{Q})$ but higher max entropy $H_\text{max}(\mathcal{P})\geq H_\text{max}(\mathcal{Q})$. The first statement holds due to $\braket{\mathcal{Q}^2}=\braket{\mathcal{P}^2}$ and $\braket{\mathcal{Q}}^2 \geq \braket{\mathcal{P}}^2=0$. The second statement follows from the concavity of the entropy function:
\begin{align*}
H_\text{max}(\mathcal{P}) &= 2\log_{2}\sum_k{\sqrt{p_{k}}} \\
			&=
                   2\log_{2}\sum_k{\sqrt{\frac{q_{k}+q_{-k}}{2}}}\\
  &\geq 2\log_2 \sum_k\left(
    \frac{1}{2}\sqrt{q_{k}}+\frac{1}{2}\sqrt{q_{-k}}\right)\\
  &=H_\text{max}(\mathcal{Q})\,.
\end{align*}
Hence, the extremal distribution is symmetric and has zero mean. 

To find the extremal distribution $\mathcal{P}$, we write the Lagrangian as:
\begin{multline*}
\textsl{L}(\mathcal{P},\alpha,\gamma) =  2\log_{2}\sum_k
{\sqrt{p_{k}}}\\
+ \frac{\alpha}{\ln 2}\left(1- \sum_{k}{p_{k}}\right)+
\frac{\gamma}{\ln 2} \left(V-\sum_{k} p_{k}\,x_k^{2}\right)\,.
\end{multline*}
$\textsl{L}$ attains a stationary point when
\begin{align*}
   &&\frac{\partial\textsl{L}}{\partial p_{k}} &= 0\\
   &\Rightarrow& \frac{1}{\sqrt{p_k}}\frac{1}{\sum_j \sqrt{p_j}} - \alpha - \gamma \,x_k^{2}&=0\\
  &\Rightarrow&\frac{1}{\sqrt p_{k}} &=  \left(\alpha + \gamma\, x_k^{2}\right)\sum_j \sqrt{p_j}\,.
\end{align*}
Multiplying both sides by $p_k$ and summing over $k$, we obtain the relation:
\begin{align*}
  \alpha + \gamma V =1\,.
\end{align*}
This together with the constraint $\frac{\partial \textsl{L}}{\partial
  \alpha}=0$ allows us to write
\begin{align*}
  p_{k} &= \frac{\frac{1}{\left(1+\gamma(x_k^2-V)\right)^2}}
            {\sum_j \frac{1}{\left(1+\gamma(x_j^2-V)\right)^2}}\,.
\end{align*}
We recognize this as a discretised version of the non-standardized Student's
t-distribution with 3 degrees of freedom and standard deviation $s$:
\[S_3(x;s) = \frac{2}{\pi s\left(1+\frac{x^2}{s^2}\right)^2}\,.
\]
When $\delta q \rightarrow 0$ and $m\, \delta q \rightarrow \infty$, we retrieve the continuous limit, $\gamma \rightarrow \frac{1}{2}$ and $s^2 \rightarrow V$. This is consistent with the known result that the Student's t-distribution are the extremal continuous distribution for $H_\text{max}$~\cite{johnson_results_2007}.

A necessary condition for the Lagrange multiplier $\gamma$ is obtained from the constraint $\frac{\partial \textsl{L}}{\partial \gamma}=0$ which gives an implicit equation:
\begin{align*}
&&\sum_k   \frac{x_k^2}{\left(1+\gamma(x_k^2-V)\right)^2}
  &=  \sum_j \frac{V}{\left(1+\gamma(x_j^2-V)\right)^2} \\
    & \Rightarrow& \sum_k\frac{x_k^2-V}{\left(1+\gamma(x_k^2-V)\right)^2} &=0\,.
\end{align*}
Numerically, we see that there can be more than one real solution for $\gamma$. The extremal $H_\text{max}$ is given by the solution that is closest to zero.

\section{Small number of bins example}
\label{sec:smalln}
\begin{figure*}[t!]
	\centering
	\includegraphics[scale=0.5]{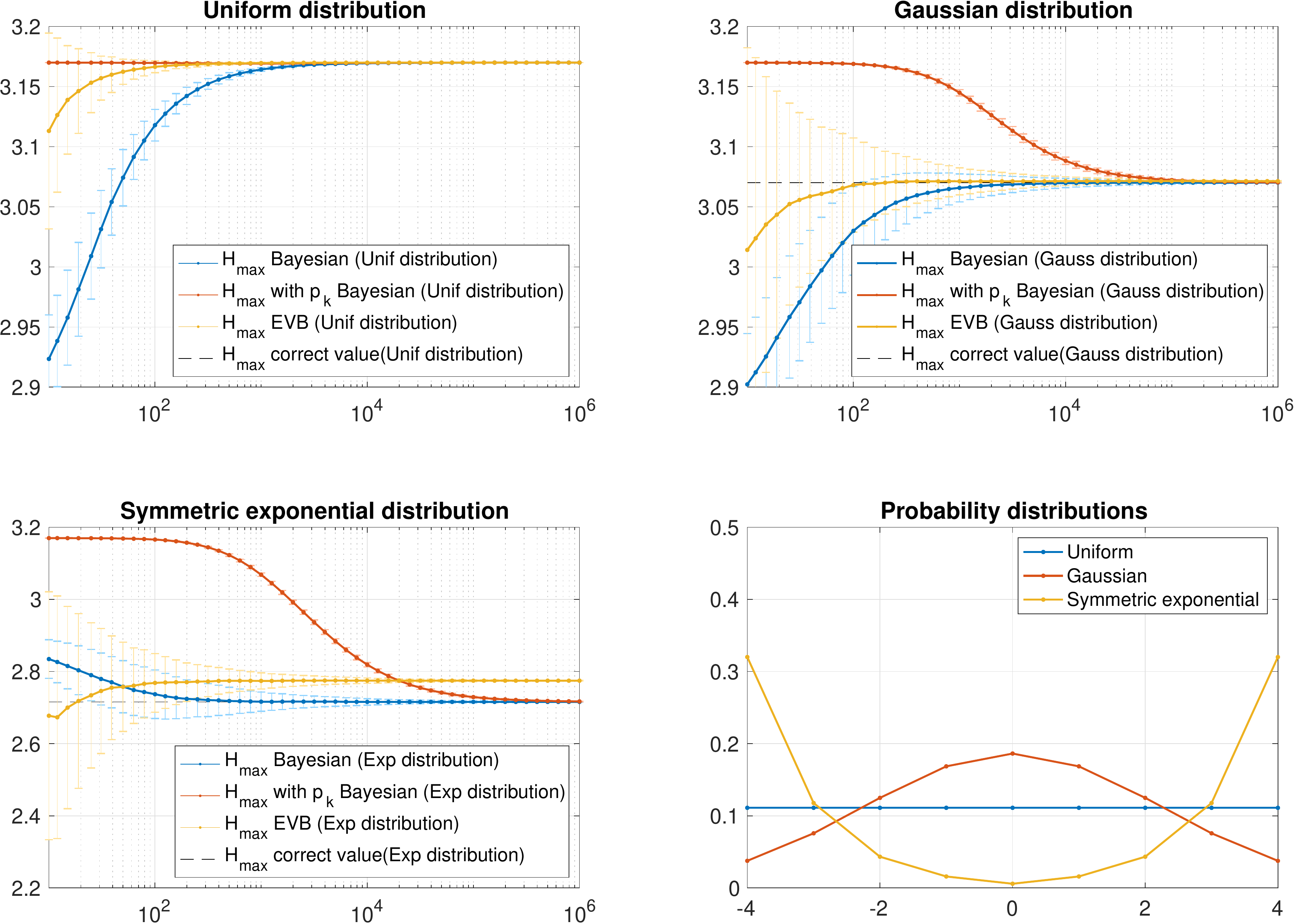}%
	\caption{\label{low_bin_comparison} Comparison of estimators
          for $H_\text{max}$ on three probability distributions with
          just nine
          bins. A negative bias on $H_\text{max}$
          translates to a positive bias on $H_\text{low}$. For the
          Bayesian estimator with peaked prior, $K$ was set to $100$.}
\end{figure*}

In this Appendix, we show that under extreme cases when the number of
bins is very small, when the number of samples are very small or when
the input state saturates the extreme bins, some of the estimator for
$H_\text{max}$ proposed in the main text may still be negatively
biased, which would lead to a positive bias on $H_\text{low}$.  To
illustrate this we considered three different distributions with only nine
bins as shown in Fig.~\ref{low_bin_comparison}. The only estimator
that shows no negative bias is the peaked prior Bayes estimator.

\end{document}